1# PON Monitoring Technique Using Single-FBG Encoders and Wavelength-to-Time Mapping

Manuel P. Fernández, Laureano A. Bulus Rossini, Pablo A. Costanzo Caso, *Member*, *IEEE*.*Abstract*—We present a novel coding-based technique for remote monitoring of passive optical networks which exploits the wavelength-to-time mapping effect of broadband short pulses in a highly dispersive medium. In this scheme, each encoder consists in a single FBG written at the same central wavelength but each one having a unique spectral bandwidth. The ultra-compactness and low-cost of the encoders allow them to be potentially integrated inside the customers' terminal. We experimentally demonstrate the feasibility of the proposed method for the supervision of current and future high-density networks.

*Index Terms*—Passive Optical Network (PON) monitoring, FBG encoders, Wavelength-to-time mapping.## I. Introduction

Passive optical networks (PONs) have been massively deployed in recent years to support the growing bandwidth demand in the last mile. From the operators' perspective, it is essential to have a remote monitoring system to improve service reliability and to reduce operational costs by timely detecting and localizing potential faults (e.g. a fiber break) [1]. The use of conventional Optical Time-Domain Reflectometry (OTDR) suffers from severe limitations when applied to PONs that contain splitters since the acquired waveform results from the superposition of backscattered power from several branches [2]. In order to individually supervise each user within the PON, fiber Bragg gratings (FBGs) at different wavelengths can be placed at the end of each branch [3]. However, this is impractical as the number of users increases, since too many wavelength channels would be required.

To overcome this limitation, several monitoring schemes based on optical code division multiplexing (OCDM) have been reported [4]-[11]. In these proposals, a passive optical encoder composed by two or more FBGs is placed at the customers' side, which generates a unique signature sequence for each user from a single monitoring probe pulse. At the central office (CO), the sum of all codes is received and the status of each branch is derived by performing a decoding

This work was partially supported by Consejo Nacional de Investigaciones Científicas y Técnicas (CONICET), Universidad Nacional de Cuyo UNCuyo), Comisión Nacional de Energía Atómica (CNEA) and Sofrecom Argentina SA. M.P.F., P.A.C.C and L.A.B.R are with Laboratorio de Investigaciones Aplicadas en Telecomunicaciones, CNEA, Bariloche 8400, Argentina. M.P.F. is fellow of CONICET (e-mail: manuel.fernandez@ib.edu.ar). P.A.C.C. is a researcher in CONICET and professor at Instituto Balseiro (email: pcostanzo@ib.edu.ar). L.A.B.R. is a professor at Instituto Balseiro and researcher in CONICET (email: lbulus@ib.edu.ar). We thank the valuable comments of Professor José Azaña from INRS, Montreal, Canada.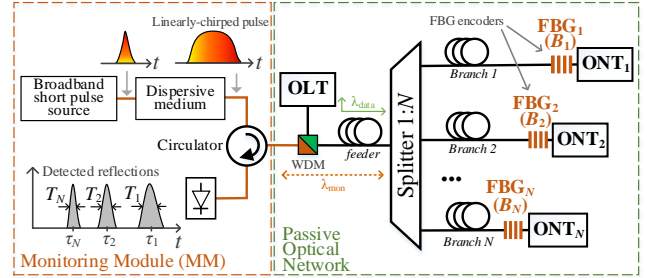

Figure 1. Schematic of the monitoring system based on single-FBG encoders.

operation over the detected signal. More recently, a monitoring scheme based on optical pulse width modulation (OPWM) was demonstrated [12], [13]. In this case, the encoder is a fiber loop composed by a FBG, a power splitter and a patch cord, which modulates the temporal width of the monitoring pulse by properly designing the patch length.

In all the previously mentioned proposals, the encoders are still relatively complex and bulky, as they require many optical components and patch cords up to several meters long, which penalizes the system's feasibility in terms of compactness and unit cost. In fact, the cost and complexity of the encoder devices should be kept as low as possible for the operator to reduce capital expenses.

In this letter, we propose a novel coding-based PON monitoring technique based on the wavelength-to-time mapping effect of broadband short pulses in a highly dispersive medium. The most remarkable feature is that the encoders are single FBGs whose spectral bandwidth serves as a signature for each user within the PON. Thus, the encoders are ultra-compact and hence their complexity, size and cost, are greatly reduced with respect to previous proposals.

## II. Monitoring principle

Figure 1 depicts the basic scheme of a PON in which the operator's optical line terminal (OLT) is connected to $N$ customer's optical network terminals (ONT) through a 1:$N$ power splitter. The monitoring system, shown in the same figure, comprises two basic elements.

*(1)* The *encoders,* each of which being a single FBG having a unique spectral bandwidth $B_i$, $i = 1, …, N$, and the same central wavelength within the U-band (1625-1675 nm), as recommended by the ITU-T L.66 Recommendation for monitoring purposes.

*(2)* A *Monitoring Module* (MM) in the CO*,* which has a structure similar to that of an OTDR, in which the probe pulse

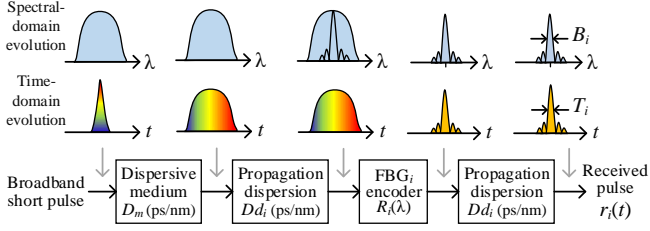

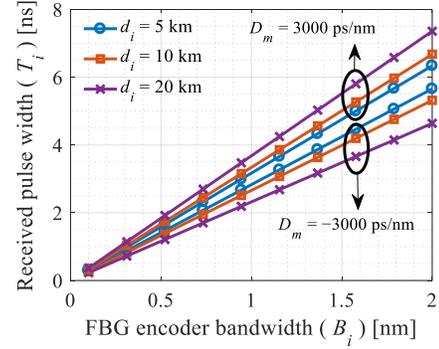
Figure 2. Diagram of the round-trip propagation of the probe pulse to the encoder $FBG_i$. Propagation losses and higher-order dispersion terms are omitted.

Figure 3. Received pulse width dependence with the FBG bandwidth. A dispersion of 3000 ps/nm is equivalent to that of ~176 km of SSMF.

is a short broadband pulse that firstly propagates through a highly chromatic dispersive medium previously to be transmitted to the PON. The linearly chirped monitoring pulse probes the encoders and the reflected pulses (i.e. codes) are detected and digitized, obtaining an OTDR-like waveform. As it will be demonstrated in the following, by measuring the temporal width of the reflected pulses $T_i$, the corresponding spectral bandwidth of the encoders $B_i$ can be deduced, and thus, each user within the PON can be identified.

*A. Mathematical background*

The round-trip propagation of the probe pulse to the $i$-th encoder, $FBG_i$, is depicted in Fig. 2. The short probe pulse first propagates through the highly dispersive medium which has an accumulated dispersion $D_m$ (in ps/nm). The resulting linearly chirped pulse probes the $FBG_i$ encoder, which is located at a distance $d_i$ from the CO, and the reflected pulse is detected back at the MM. Therefore, the overall accumulated chromatic dispersion is given by

$$D_i = D_m + 2Dd_i \quad [\text{ps/nm}], \quad (1)$$

where $D$ is the dispersion coefficient of the PON fiber (in ps/nm.km). If the total dispersion is sufficiently high, such as the temporal Fraunhofer condition is accomplished, the spectral content of the pulse is linearly mapped into its temporal waveform with a wavelength-time scale determined by the dispersion, as $\lambda = t / D_i$ [14], [15]. This effect is known in the literature as dispersive Fourier transformation or wavelength-to-time mapping. Consequently, if the spectrum of the input pulse is $X(\lambda)$, the resulting temporal waveform would be $x(t) \propto X(t / D_i)$.

The $FBG_i$ encoder is described by the reflectivity transfer function $R_i(\lambda)$, and thus the detected pulse corresponding to the $FBG_i$ encoder can be expressed as

$$r_i(t) \propto R_i\left(\lambda = \frac{t - \tau_i}{D_i}\right). \quad (2)$$

where it was reasonably assumed that the FBG encoder has negligible chirp compared to $D_i$, and that the probe pulse spectrum is approximately flat within the encoder's bandwidth. According to Eq. (2), the detected pulse has a shape identical to that of the $FBG_i$ reflectivity spectrum, with a wavelength-time scale given by the accumulated dispersion $D_i$. Moreover, the pulse is acquired with a delay $\tau_i$, which is associated to the encoder's distance through the relation $\tau_i = 2d_i/v_g$, with $v_g$ being the group velocity at the monitoring wavelength. Therefore, if the function $R_i(\lambda)$ has a spectral full-width at half maximum (FWHM) $B_i$, then the detected pulse has an equivalent temporal FWHM $T_i$ given by

$$T_i = |D_i| B_i = |D_m + 2Dd_i| B_i. \quad (3)$$

Thus, the OTDR-like waveform acquired by the MM is sent to a network recognition algorithm that measures the temporal FWHM of all the pulses and identifies each encoder through its bandwidth using Eq. (3), since the parameters $D_m$ and $D$ are previously known, and $d_i$ is derived from the round-trip propagation delay $\tau_i$.

Figure 3 shows the relation between the detected pulse width and the encoder's bandwidth considering different system parameters. We contemplate a dispersive medium in the MM having a dispersion parameter of 3000 ps/nm with both negative and positive signs. Additionally, we consider distances typical of current PON deployments of 5 km, 10 km, and 20 km. The PON fiber is assumed to be a standard single-mode fiber (SSMF) with $D = 17$ ps/nm.km. It is seen that FBG bandwidths of a few hundreds of picometers yield pulses of a few nanoseconds. As it is expected from Eq. (3), the slope of the function depends on the overall dispersion, and it increases with the distance to the encoder $d_i$ in the case that $D_m$ has the same sign that $D$, while it decreases if they have opposite sign.

*B. Encoders design*

It is necessary to establish a design criterion to assign the spectral bandwidth $B_i$ to each encoder such that they can be univocally identified from the measured pulse width, even under the most pessimistic detection conditions. Let us assume that the detection system is capable of discerning pulse widths with a resolution $\delta T$. Let us also define the minimum possible accumulated dispersion in the system, $D_{\min} = \min\{|D_i|\}$. By considering this, we can establish that the minimum difference between two FBGs spectral bandwidths must be $\delta B = \delta T / D_{\min}$. Thus, the encoders' bandwidth can be assigned according to the following rule

$$B_i = B_1 + (i-1)\delta B \quad i = 1,...,N \quad (4)$$

where $B_1$ is the minimum bandwidth, and it is determined by the minimum measurable pulse width.

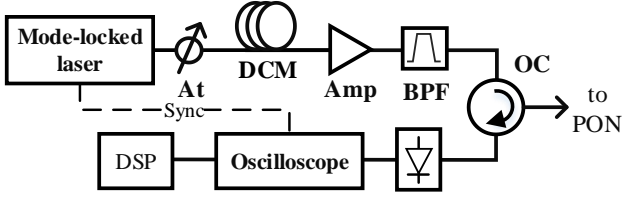

Figure 4. Experimental setup of the monitoring module. DCM: Dispersion Compensation Module, At: Attenuator, Amp: Amplifier, BPF: Band-pass filter, OC: Optical Circulator, DSP: Digital Signal Processing.

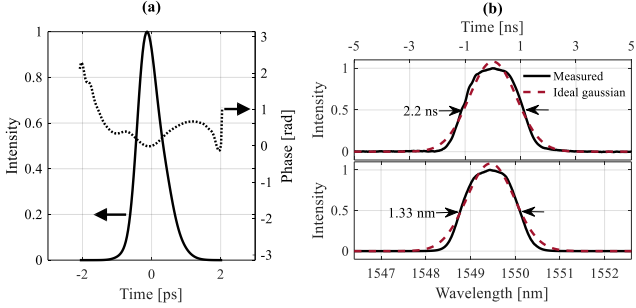

Figure 5. (a) Probe pulse at the output of the laser source. (b) Wavelength-to-time mapping of the probe pulse at the output of the DCM.

On one hand, a high dispersion parameter increases the system's resolution, and therefore the ability to discern between two close spectral bandwidths, as it leads to broader detected pulses. At the same time, it will increase the interference probability, which arises when two or more pulses from different encoders are temporally overlapped in the acquired signal. In the case that codes are overlapped in time, a maximum likelihood sequence estimation algorithm can be used to identify the different codes within the acquired waveform. However, a high number of interferences can yield a prohibitive decoding time [13].

Consequently, the selection of the dispersion value should result from a trade-off between a desired sensitivity and an interference probability limit. For example, 64 codes can be generated between $B_1 = 200$ pm and $B_{64} = 3.35$ nm with steps of $\delta B = 50$ pm. In such a case, if the temporal resolution is $\delta T = 250$ ps, it would be needed a minimum dispersion of $D_{min} = 5000$ ps/nm to ensure that all the different bandwidths can be temporally resolved. With these previously assumed parameters, the maximum received pulse width corresponds to $FBG_{64}$ and it is found to be $T_{64} = 16.75$ ns. In such a case, the customers can be located at distances as close as 1.67 m without having interference, i.e. temporal overlap.

From a manufacturing perspective, the spectral bandwidth of a uniform FBG can be adjusted during the writing process by properly choosing the effective index variation or the grating length. For instance, ultra-short FBGs can reach bandwidths of up to several nanometers [16].

### III. Proof-of-Concept Experiments

We deployed a test-bed PON based on the schematic of Fig. 1. Although here we used the C-band for the proof-of-concept experiments, the setup can be equally extended to the monitoring U-band. The experimental setup of the MM is show in Fig. 4. As short pulse generator we used a mode-locked laser emitting at 1550 nm with a repetition period higher than the maximum round-trip time in the PON. A dispersion compensation module (DCM), consisting of a single-mode fiber with accumulated dispersion at the monitoring band $D_m = -1659$ ps/nm is used as the highly dispersive medium. The resulting pulse is amplified, band-pass filtered and transmitted to the PON. The return signal is detected using a 5 GHz-bandwidth photodiode and digitized using a real-time oscilloscope with a bandwidth of 4 GHz and a sampling rate of 20 Gsps.

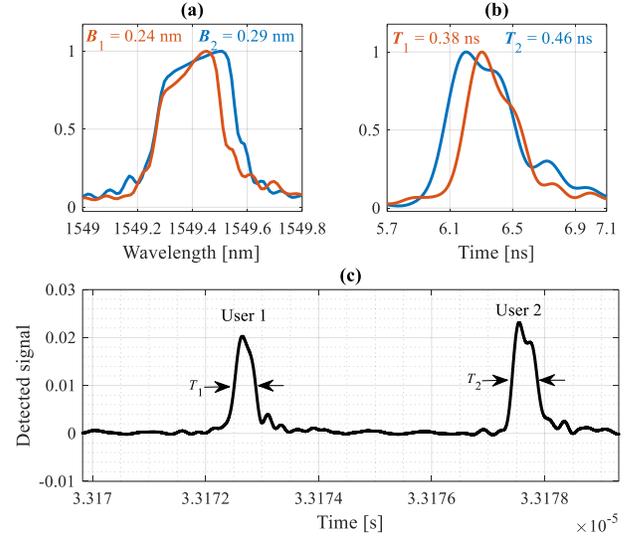

Figure 6. (a) Spectrum of the encoders $FBG_1$ and $FBG_2$. (b) Detected codes in the time-domain. (c) Example of acquired signal when two encoders are located at 3.3 km from the monitoring module.

The input probe pulse was first characterized by measuring its amplitude and phase, as depicted in Fig. 5(a). The pulse width at the DCM input is measured to be 0.81 ps, and thus the dispersive medium largely accomplishes with the temporal Fraunhofer condition [14]. Figure 5(b) shows the measured linearly chirped pulse at the DCM output in both the time and wavelength domains. The pulse presents a Gaussian profile in which the spectrum is mapped into a temporally inverted waveform, due to the negative sign of the dispersion parameter. The measured temporal and spectral FWHM are 2.2 ns and 1.33 nm, respectively.

In order to demonstrate the feasibility of the method, we used two FBG encoders, $FBG_1$ and $FBG_2$. Their transfer functions were first individually measured using an optical spectrum analyzer and they are shown in Fig. 6(a), where it is seen that both encoders share the same wavelength channel, while presenting different bandwidths (FWHM) of $B_1 = 0.24$ nm and $B_2 = 0.29$ nm, yielding a bandwidth step of $\delta B = 50$ pm. The individually detected pulses reflected from the encoders in the time-domain is shown in Fig. 6(b). It is clearly appreciated how the encoders generate reflected pulses with different temporal width, and hence each single FBG generates a unique temporal signature. In this case, the pulses FWHM are measured to be $T_1 = 0.38$ ns and $T_2 = 0.46$ ns.

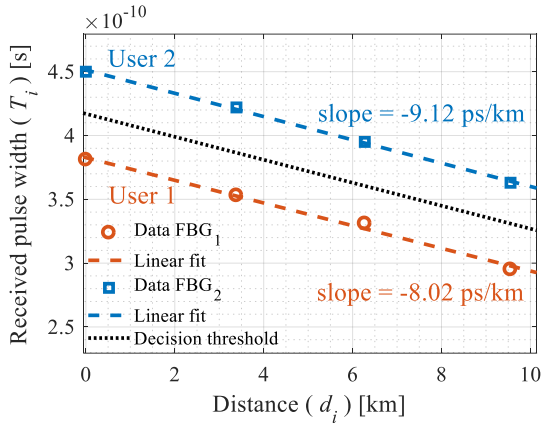

Figure 7. Received pulse width versus the distance to the FBG encoder.

Figure 6(c) shows an example of the acquired waveform when encoders $FBG_1$ and $FBG_2$ are located after a power splitter at a distance of 3.3 km from the MM, i.e. a round-trip time of approximately 330 µs. In the MM, a simple algorithm identifies the reflections within the acquired waveform and measures their FWHM. The algorithm is straightforward to implement and requires extremely low computational cost. In this example, the distance difference between the encoders' locations is only 48 cm, and both users can be individually identified through the measured reflections.

Finally, we experimentally analyzed the dependence of the detected pulse width with the distance to the encoder. Specifically, as it is shown in Fig. 7, we measured the temporal FWHM of the reflected pulses for the encoders $FBG_1$ and $FBG_2$ when considering distances to the encoders $d_i$ up to 10 km. It is clear how the different single-FBG encoders generate a unique signature (i.e. a unique pulse width) for each user for the entire distance range. The linear fit of the measured data has a negative slope since the PON fiber has opposite sign to that of the DCM. The slopes of the fitting functions are found to be −9.12 ps/km and −8.02 ps/km for encoders $FBG_1$ and $FBG_2$, respectively. As expected from Eq. (3), higher encoder's bandwidth lead to a higher slope. In the recognition algorithm, decision thresholds are set in the middle point between the function $T_i(z_i)$ of adjacent users. For instance, in the example of Fig. 7, the decision threshold to discern between $FBG_1$ and $FBG_2$ is set to be in the middle point of the two linear fitting functions.

The use of a dispersive medium with a higher dispersion, such as a dispersive fiber or chirped gratings [15], will yield to broader detected pulses, i.e. a higher offset in the linear functions of Fig. 7, which will in turn allow for the use of slower acquisition electronics.

## IV. CONCLUSIONS AND FUTURE WORK

We have proposed a novel coding-based PON monitoring system. In this method, each encoder consists of a single FBG having a unique bandwidth, which makes it the simplest and more compact encoder structure proposed up to date. The FBGs are interrogated using an OTDR-like device that emits probe pulses in the far-field temporal Fraunhofer regime. Thus, the spectral bandwidth of the FBG encoders, which serve as a signature for each user, can be derived from the measured pulse width due to the wavelength-to-time mapping property. We have presented the mathematical background and demonstrated the system feasibility through proof-of-concepts experiments. We have shown that users located as close as a few centimeters can be individually identified.

Future work includes extending the proposed method to a multi-wavelength approach, in which several wavelength channels are used. This would enormously increase the network capacity and reduce the interference probability.

REFERENCES

[1] M. Esmail and H. Fathallah, "Physical Layer Monitoring Techniques for TDM-Passive Optical Networks: A Survey", *IEEE Commun. Surveys and Tutorials*, vol. 15, no. 2, pp. 943-958, 2013.
[2] M. P. Fernández, L. A. Bulus Rossini, J. P. Pascual and P. A. Costanzo Caso, "Enhanced fault characterization by using a conventional OTDR and DSP techniques", *Opt. Express*, vol. 26, no. 21, p. 27127, 2018.
[3] K. Tanaka, M. Tateda and Y. Inoue, "Measuring the individual attenuation distribution of passive branched optical networks", *IEEE Photon. Technol. Lett.*, vol. 8, no. 7, pp. 915-917, 1996.
[4] H. Fathallah, M. Rad and L. Rusch, "PON Monitoring: Periodic Encoders With Low Capital and Operational Cost", *IEEE Photon. Technol. Lett.*, vol. 20, no. 24, pp. 2039-2041, 2008.
[5] M. Rad, H. Fathallah, M. Maier, L. Rusch and M. Uysal, "A Novel Pulse-Positioned Coding Scheme for Fiber Fault Monitoring of a PON", *IEEE Commun. Lett.*, vol. 15, no. 9, pp. 1007-1009, 2011.
[6] M. P. Fernandez, P. A. Costanzo Caso and L. A. Bulus Rossini, "False Detections in an Optical Coding-Based PON Monitoring Scheme", *IEEE Photon. Technol. Lett.*, vol. 29, no. 10, pp. 802-805, 2017.
[7] M. P. Fernández, P. A. Costanzo Caso and L. A. Bulus Rossini, "Design and performance evaluation of an optical coding scheme for PON monitoring," in *XVI Workshop on Information Processing and Control*, Cordoba, 2015.
[8] X. Zhou, F. Zhang and X. Sun, "Centralized PON Monitoring Scheme Based on Optical Coding," *IEEE Photon. Technol. Lett.*, vol. 25, no. 9, pp. 795-797, May 2013.
[9] L. Baudzus and P. M. Krummrich, "CDM-Based lambda-OTDR for Rapid TDM- and WDM-PON Monitoring," *IEEE Photon. Technol. Lett.*, vol. 26, no. 12, pp. 1203-1206, June 2014.
[10] M. Thollabandi, X. Cheng and Y. Yeo, "Encoded Probing Technique for Detection of the Faulty Branch in TDM-PON", *IEEE Photon. Technol. Lett.*, vol. 24, no. 18, pp. 1610-1613, 2012.
[11] X. Zhang, S. Chen, F. Lu, X. Zhao, M. Zhu and X. Sun, "Remote Coding Scheme Using Cascaded Encoder for PON Monitoring," *IEEE Photon. Technol. Lett.*, vol. 28, no. 20, pp. 2183-2186, 15 Oct.15, 2016.
[12] X. Zhang, H. Guo, X. Jia and Q. Liao, "Fault localization using dual pulse widths for PON monitoring system", *Opt. Laser Technol.*, vol. 106, pp. 113-118, 2018.
[13] X. Zhang and X. Sun, "Optical pulse width modulation based TDM-PON monitoring with asymmetric loop in ONUs", *Sci. Rep.*, vol. 8, no. 1, 2018.
[14] K. Goda and B. Jalali, "Dispersive Fourier transformation for fast continuous single-shot measurements", *Nature Photon.*, vol. 7, no. 2, pp. 102-112, 2013.
[15] J. Azana and M. Muriel, "Real-time optical spectrum analysis based on the time-space duality in chirped fiber gratings", *IEEE J. Quantum Electron.*, vol. 36, no. 5, pp. 517-526, 2000.
[16] R. Cheng, L. Xia, C. Sima, Y. Ran, J. Rohollahnejad, J. Zhou, Y. Wen, and C. Yu, "Ultra-short FBG based distributed sensing using shifted optical Gaussian filters and microwave-network analysis," *Opt. Express* vol. 24, no.3, pp. 2466-2484, 2016.